# Thematic Identification of “Little Science”: Trends in Portuguese IS&LS Literature by Controlled Vocabulary and Co-Word Analysis

Silvana Roque de Oliveira 1 Catarina Moreira 2 José Borbinha 3 María Ángeles Zulueta Garcia 4

1 Facultad de Documentación, Universidad de Alcalá, Spain
2 Instituto Superior Técnico / INESC-ID, Portugal
3 Instituto Superior Técnico / INESC-ID, Portugal
4 Facultad de Documentación, Universidad de Alcalá, Spain

**Abstract:** This study presents an overview of IS&LS thematic trends in Portugal between 2001 and 2012. The results were obtained by means of an analysis, using expeditious qualitative and quantitative techniques, of the bibliographic records of proceedings papers identified during this period. These records were processed using two techniques: a manual subject classification and an automated co-word analysis of the Author-Assigned Keywords. From this we designed cluster and co-occurrence maps, using the VOSviewer and the Pajek software packages.
The results indicated an accentuated dynamism in the thematic evolution of this documental corpus, apart from revealing a significant difference among the themes transmitted in nationally and internationally visible production.

…………………………………………………………………………………………………………

## 1. Introduction

This study presents the results of a thematic analysis of the Portuguese IS&LS community.
Becoming better acquainted with the subjects that are of greater interest for authors of a given community makes it possible to develop the epistemological perception which each discipline and the respective scientific community have of themselves. Moreover, this could also be useful to guide the development of editorial policies and library collections, which could thus more easily satisfy information needs in each knowledge domain.

While examining the area of IS&LS it is necessary to consider two aspects which distinguish it from other areas of scientific research. On the one hand, its eminently interdisciplinary nature. On the other, the fact that it is an area which began by being a profession (Hjørland, 2000, López Cozar, 2002; Silva et al., 2002), even today evidencing different levels of maturity in terms of its scientific affirmation. In the case of Portugal, it can be considered to be "little science", using the epithet coined by Price (1963). In this regard, Koehler (2001) stated: "One way to mark the transition from little science to big for any given discipline is to document the change from practitioner to scientist among its participants". In Portugal exclusive dedication to research is undoubtedly still at a nascent stage (Calixto, 2008). Its history is atypical in the context of the western world even in terms of training, the first graduate programme having been created only in 2001. Until then there were only post-graduate training courses, essentially aimed at imparting professional skills, for librarians, archivists and documental specialists (Ribeiro, 2006; Pinto, 2008). This situation resulted in the absence of Portuguese publications in leading international databases. This fact has hindered an effective knowledge of Portuguese production standards. Among the rare bibliometric studies of this area especially worthy of note is Cerqueira & Silva (2007), which analysed a Portuguese publication compiled directly from primary sources, and Moya-Anegón & Herrero-Solana (2002), Herrero-Solana & Liberatore, (2008), Olmeda Gómez, Perianes-Rodríguez & Ovalle-Perandones (2008), cases reporting analyses based solely on data available at the Web of Science (WoS). All these studies have emphasised the indicators of production, authors and visibility to the detriment of thematic trends, a gap that this study seeks to overcome.

## 2. Research Goals

The objective of the task at hand was to produce a profile of IS&LS thematic trends in Portugal, developed in a corpus of proceedings papers between 2001 and 2012. In the context of a vaster project of a bibliometric analysis of Portuguese IS&LS production, a very small part of which is presented herein, the fact of being directly in contact with the texts was viewed as an opportunity. This circumstance made it possible to apply two distinct thematic analysis methods in a complementary manner: one being qualitative, involving a manual classification based on a list of controlled vocabulary; the other being quantitative, using co-word analysis. The results were made visible by means of two freeware software packages used for information visualisation, Pajek 3.11(2013) and VOSviewer 1.5.4 (2013).

## 3. Literature Review

The bibliographic classification is based on the use of sets of concepts which provide a structured description of a given thematic universe, thus constituting a pre-understanding of the areas to which they are applied, in a deductive approach. While examining IS&LS research, López Cózar (2002) reviewed the contributions of more than twenty studies dedicated to a thematic profile, published between 1976 and 2000. A comparison reveals a manifest preference for the manual application of classifications constructed a priori. An inductive approach could be an alternative, creating the terms of classification according to what is most pertinent with regard to the documentation being analysed. Although his results were undoubtedly valid, authors considered them to be insufficiently streamlined in order to establish comparisons (Atkins, 1988; Koufogiannakis, Slater & Crumley 2004), thus suggesting the application of quantitative methods for this purpose, namely resorting to the co-word analysis of titles or keywords. Nowadays, the discussion regarding tools for knowledge organisation and information retrieval always contemplates the challenges raised by the descriptive potential of natural language, while continuing to recognise the validity of controlled languages (Freire, Borbinha and Calado, 2011; Schwing, McCutcheon & Maurer, 2012; Engelson, 2013). There is hence a growing tendency towards the complementary use of these techniques. If natural language lacks normalisation, controlled languages also lack the context and the uniqueness that the former usually ensures.

The use of thesauri in the field of IS&LS has already been compared with co-word analysis applied to Author-Assigned Keywords. Given the variance with natural language, co-word analysis proved to be very effective in capturing the dynamism of the thematic evolution in this area, an interesting complementarity thus emerging between the two approaches (Ding & Foo, 2001). Co-word analysis is a content analysis technique where the objective is to identify and track the evolution of issues that are of most interest to scientists, based on relationships established between pairs of words (He, 1999). Unlike controlled languages, co-word analysis works directly with the raw material of thematic construction, the words used by the authors themselves and their different weighting, using automatic forms of observation, in the conceptual framework of an analysis of social networks (Callon et al., 1983; Yi & Choi, 2012). This method revolutionised the thematic knowledge which can be obtained from large documental masses, making it possible to prepare scientific maps (Moya-Anégon et al., 2004). Recently studies have been developed about IS&LS using co-word analysis, based on different types of documents and using keywords or the titles of publications (Milojevic et al. 2011; Liu et al., 2012; Zong, et al., 2013).

## 4. Material and Methods

To identify the main topics developed by this discipline in Portugal, we chose proceedings papers as our data source. Conferences have been playing an increasingly relevant role in scholarly communication, because they promote the presentation of the latest issues in each field of knowledge and are privileged spaces for scientists to expand personal social networks (González-Albo & Bordons, 2011). An analysis of proceedings papers is even more pertinent in areas with an accentuated technical or professional profile (López-Cózar, 2002; Glänzel, Schlemmer & Schubert, 2006).

### 4.1. Data set and Procedures

Of Portuguese conferences in this area, the Congresso Nacional de Bibliotecários, Arquivistas e Documentalistas (hereinafter BAD), organised by the eponymous professional association, is the only conference which regularly publishes proceedings and follows editorial norms which enable a comprehensive bibliometric analysis. Thus, we decided to compile all the 276 papers presented in five of its eleven editions in which the proceedings were published in a digital format, between 2001 and 2012 (APBAD, 2013). With regard to internationally visible Portuguese production, we decided to retrieve the papers indexed in the WoS . After the necessary filtering of the data, we retained 156 proceedings papers. In total, 432 proceedings papers were analysed.

### 4.2. Controlled Vocabulary Classification

We developed a manual classification for the 432 documents, using the scheme proposed by Järvelin & Vakkari (1990 and 1993) for IS&LS, corroborated by López-Cózar (2002) and used again in a report, for various national cases, by Rochester & Vakkari (2004). Each document was attributed only one class. As one of the benefits is the comparability with previous studies, we stayed as close as possible to the original list of nine classes. However, in the process of comparing the documents retrieved by the WoS, it proved necessary to add a tenth class for "Other disciplines". This situation can be explained by the fact that the WoS thematic categories classify journals and not the articles individually, thus retrieving data from a variety of disciplines whose boundaries are very difficult to define (Gómez et al., 1996), especially in an interdisciplinary field such as IS&LS.

Since the documental corpus also included contributions from the archival sector, some classes were adjusted accordingly, opting for broader designations such as "Institutional Evolution" and "Edition & Document Evolution". Our final list had the following terms: Profession; Edition & Document Evolution; Information Services; Information Organization & Retrieval; Information Search; Scientific Communication; Other Topics; Other Disciplines.

Finally, given that Portuguese production is essentially applied, we decided to develop a second ad hoc list of 9 terms in order to retrieve the different types of information services associated with the documental corpus, namely: Archives, Libraries, Documentation Centres; Archives&Lib&DC; Organisations; Databases; Internet; Other Institutions; Theoretical Approach. The data processing was limited to considering the absolute and relative frequencies.

### 4.3. Co-Word Analysis and Information Visualisation

Since we noticed that many titles were too emphatic for the objectives of this type of research, Author-Assigned Keywords seemed to be the most suitable analysis unit given the traits of the documentation (Milojevic et al., 2011). In light of a far smaller set of data than usual in such studies – we are in a context of "little science" –, we tested a different strategy from the one developed in previous studies (Vargas-Quesada et al., 2007; Zulueta et al., 2011; Cantos-Mateos et al., 2012), which enabled a rapid visualisation of data without it having been necessary to resort to algorithms to select subgraphs.

In an initial phase, we began by automatically translating the keywords into English, we corrected the semantic options according to the vocabulary recognised in IS&LS and we harmonised the concepts which had two exactly equal forms of meanings, thus obtaining 877 descriptors with a total of 1474 occurrences.

In order to analyse co-occurrence and to visualise the relationships between descriptors, it was necessary to organise the information of the proceedings papers into a relational database in Microsft SQL Server 2012. Based on this relational model, the co-occurrence of the descriptors was counted automatically. Given that it was impossible to obtain an image legible to the naked eye with all the 877 existing descriptors, a criterion was established to retrieve only descriptors with 5 or more occurrences. The sample to be analysed was thus reduced to the 46 most significant descriptors, with a total of 388 occurrences, comprising 26% of the universe being studied.

For the graphic drawing of the spatial descriptors we used the Pajek and VOSviewer software packages (Van Eck & Waltman, 2010).

Pajek enables a more spatial analysis of the graphs, while VOSviewer enables an analysis of the graphs of the clusters of descriptors with similar characteristics. A graph can be defined as a set of vertexes interlinked by means of a set of edges. In order to be able to generate the graphs, Pajek needs the information of the vertexes as well as of the edges. The information regarding the edges, i.e. regarding the relations between the descriptors, was obtained through the database. Each edge is associated with a weighting which corresponds to the number of times that a word co-occurs with another (Greiff (1998). Having all this information (Fig. 1 shows an example of the results of a query to our database), it was possible to generate the graph and analyse it using Pajek.

Fig. 1 – Results of a query to identify edges

| Keyword1 | Keyword2 | weight |
|---|---|---|
| academic libraries | citizenship | 1 |
| academic libraries | collaboration | 3 |
| academic libraries | digital libraries | 1 |
| academic libraries | evaluation | 2 |

In order to obtain information about clusters of descriptors, we used the features of VOSviewer (Fig. 2). This software automatically calculates the most significant groupings. For our analysis, the descriptors were grouped according to their themes. Descriptors of thematically related topics tend to be closer to each other, forming a cluster. We then exported this information and incorporated it into Pajek, selecting the Kamada-Kawai algorithm to visualise it. We did not limit ourselves to using VOSviewer since this programme does not allow a clear visualisation of the relationships between the various descriptors.

## 5. Results

In terms of classification with controlled vocabulary, this article will limit itself to describing the most expressive values of the first thematic list here, while acknowledging the interpretative potential of cross-referencing the two lists. The thematic evolution, distributed over two equal chronological intervals (2001-2006 and 2007-2012) for the set of documents retrieved in BAD and WoS, reveals that the area has remained quite stable, with a greater prevalence of "Other disciplines" (26% and 31%, respectively), followed by the area of "Information Services" (20%), with a noteworthy growth in the second time period (29%). On the contrary, a significant reduction in interest for questions related to "IS&LS Education" can be noted from one period to another, declining from third place (14%) to fifth place (4%). All the other areas have values below 10%, in both chronological periods, with the exception of "Information Search", which rose from 11% to 14%. However, these general values are subject to a bias, since they mix production with national and international visibility. If instead of a chronological period a distinction was made on the basis of the source of the data – BAD and WoS –, analysing the entire time frame (2001-2012), it is evident that the strong presence of "Other disciplines" is due to 78% of the documents indexed in the WoS. The second most prevalent theme in the WoS is "Information Search", with just 12%, all the other eight themes being below 3.5%.

Moving on to an interpretation of the Author-Assigned Keywords, it is possible to note that of the 46 most significant descriptors, the top ten with the highest total frequency are: public libraries (24), academic libraries (20), Portugal (16), Information and Communication Technologies (15), digital libraries (14), information literacy (13), training (13), interoperability (12), libraries (12) and Information Science (11). Let us see what conclusions can be inferred by reading the maps which illustrate the co-word analysis.

**Fig. 2** – Label view of keywords from BAD and WoS (2001-2012) - VOSviewer 1.5.4

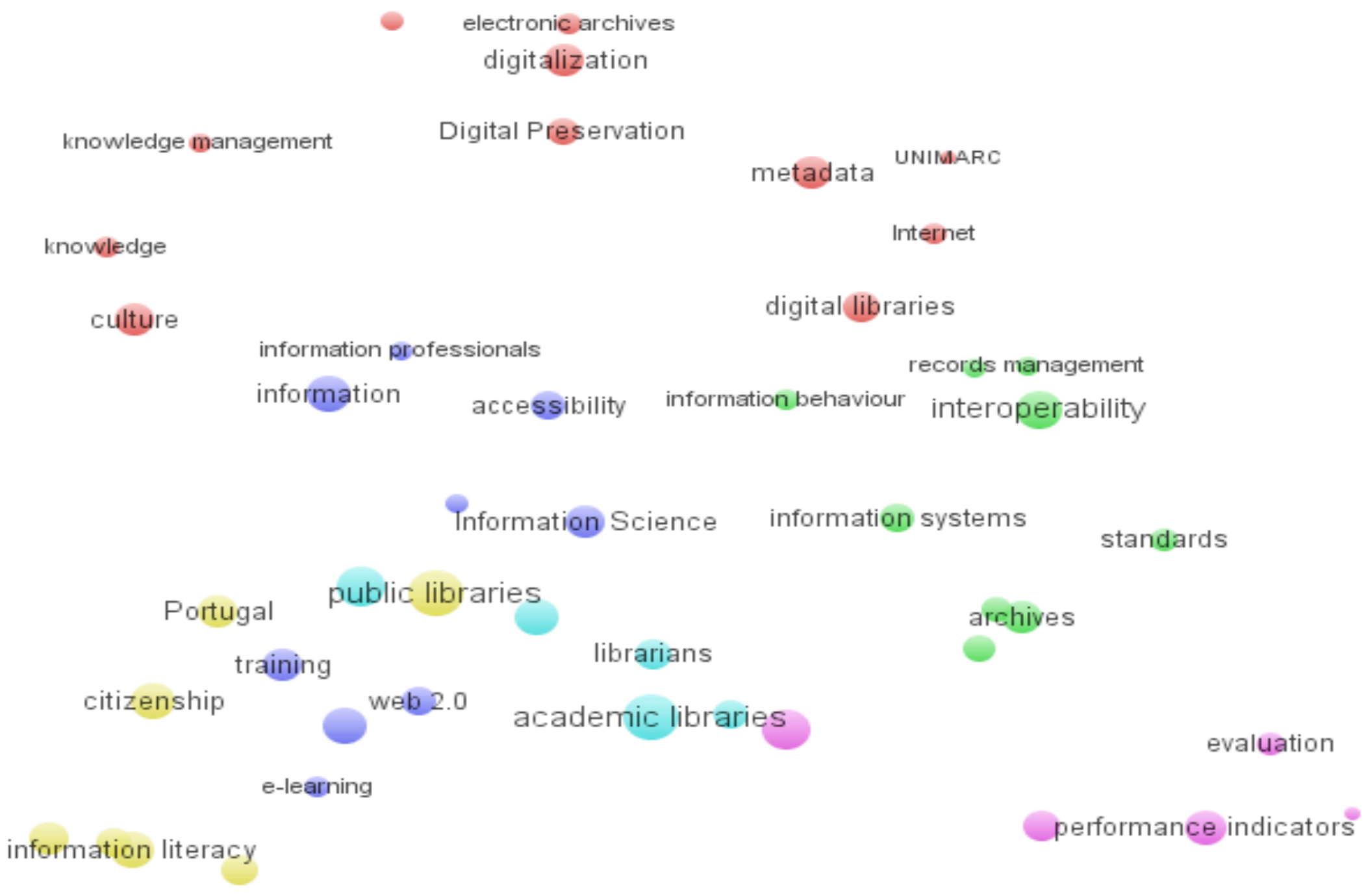

Fig. 2 identifies six clusters with different colours: Culture and Digital Preservation (11 items); Interoperability among Information Systems (9 items); Information and Training (9 items); Public

Libraries in Portugal (7 items); Performance Indicators and Evaluation (5 items) and Academic Libraries (5 items).

Figs. 3 and 4 depict a network of the co-occurrence of keywords shown in two stages. For a better interpretation it is necessary to keep in mind that the size of each node (vertex representing a descriptor) is proportional to the number of documents in which the said descriptor occurs; the thickness of the connections (edges) is proportional to the co-occurrence of the descriptors they unite; finally, the different colours continue to represent belonging to the same thematic cluster, since the maps were imported from VOSviewer.

**Fig. 3** – Co-word analysis of keywords from BAD and WoS (2001-2006) – Pajek 3.11

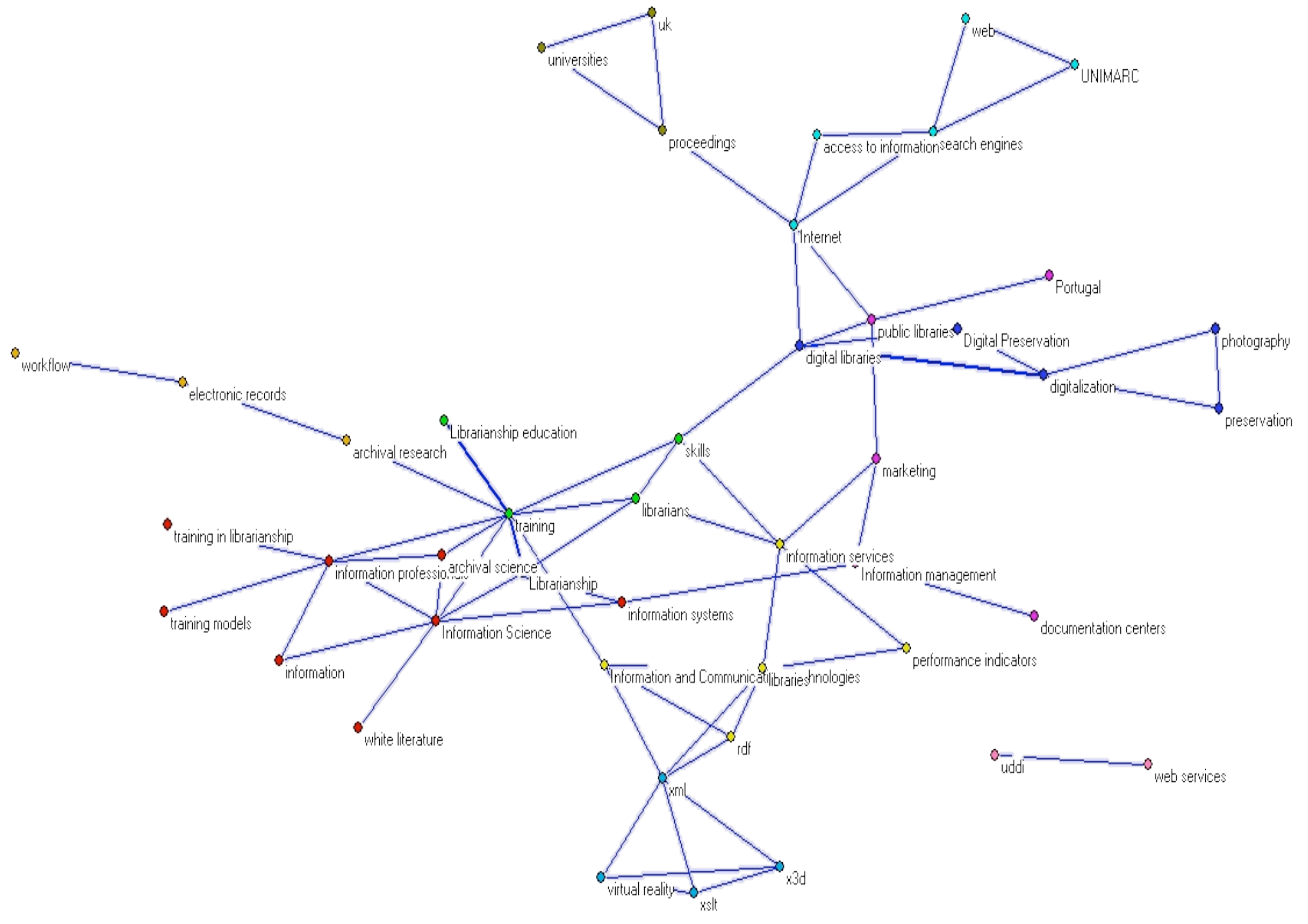

**Fig.4**– Co-word analysis of keywords from BAD and WoS (2006-2012) – Pajek 3.11

Figs. 3 and 4 are even more significant if analysed together. In truth, in the first chronological interval, IS&LS appears as an area which is not very dense, i.e., the connections are thinner, which shows that there are fewer relationships between the different themes, in addition to a lower degree of closeness – the network is widely spread. This is also due to the fact that there are far fewer documents in absolute terms (95 as compared to 337 documents). The nodes with the most relationships are “training” and “information services”, which is in keeping with what was observed by means of the manual classification. It is curious to note that "Portugal" has only one connection, to "public libraries", which suggests that the discipline contemplated the country greatly on the basis of this type of libraries, due to the prominence of the National Network of Public Libraries. In the second interval (Fig. 4), it is evident to the naked eye that the degrees of centrality, density and closeness (Vargas-Quesada et al., 2007) are much denser, the result of a far more productive and intricate discipline. New themes appear, which are easier to identify when analysing natural language. The most prominent nodes, owing to their centrality and intense closeness are "academic libraries" and "information literacy". Also worthy of note is the fact that "Portugal" is linked to many other themes, some of them very characteristic of the new national topics of interest: "open access", "repositories" and "knowledge creation".

## 6. Comparative Approach and Conclusions

In the context of research where direct contact with primary sources is a humanly feasible task, the two methods selected benefit from being used in a complementary manner and can even contribute towards a reciprocal validation of the results. The extremely brief summary presented herein clearly reveals the different level of detail which both thematic identification methods can offer, from a very general perspective of the classification to the level of detail of the co-word analysis. The two methods have far more potential than it has been possible to demonstrate here and both require a lot from researchers. Notwithstanding the strict weighting of time spent, where the advantage of co-word analysis is obvious, it would be opportune to place things in relative perspective, keeping in mind that, unlike classification, a traditional area in the competences of a librarian, co-word analysis implies a renewed investment in training, if not multidisciplinary teams such as the one which prepared this study.

The perception of discordant elements, in both methods, can also jeopardise the uniformity of the choice of the set of documents being studied (Healey, Rithman & Hoch, 1986). In the present case, there was a perceptible difference between the profile of production with national visibility (BAD) and international visibility (WoS), which makes it advisable to study the two sets separately, so that a comparison can provide useful conclusions.

It is also important to mention that the well known "indexer effect", intensely discussed during the early stages of co-word analysis (Whittaker, 1989), is present in both methods. This is because the classification with controlled vocabulary was applied by a single indexer, with all the associated idiosyncrasies as well as because the keywords which served as the basis for the analysis, chosen by diverse authors, multiply this effect in terms of the respective number of authors. The keywords can reveal the emphasis that each author wished to give to the theme more than the effective development which the theme received in the text.

As for the flexibility of the data analysis, although complementary, it is clear that the thematic classification can become obsolete more easily. On the contrary, co-word analysis is very effective at detecting quicker changes in the themes which appear, as is clear in this case study. In truth, the selected classification list proved to be more opaque in terms of uncovering some more innovative characteristics in the context of IS&LS in Portugal, with regard to the digital world or free access to information.

## Acknowledgements

This study was conducted within the scope of a Ph.D. scholarship awarded to the first author by the Fundação para a Ciência e Tecnologia (Portugal), Ref. SFRH / BD / 46188 / 2008. The authors would like to thank Gisela Cantos-Mateos for clarifications provided, as well as Roopanjali Roy for the final translation.